\documentclass[aps,prl,twocolumn,groupedaddress,showpacs]{revtex4}

\usepackage{amsmath}
\usepackage{amssymb}
\usepackage{graphicx}

\newcommand{\bel}[1]{\begin{equation}\label{#1}}
\newcommand{\bal}[1]{\begin{eqnarray}\label{#1}}

\newcommand{\imag}{\textrm{i}}

\begin{document}


\title{Damped collective motion of isolated many body systems
within a variational approach to functional integrals}


\author{Christian Rummel}
\author{Helmut Hofmann}
\affiliation{Physik-Department der Technischen Universit\"at
  M\"unchen, D-85747 Garching, Germany}


\date{\today}

\begin{abstract}
Two improvements with respect to previous formulations are
presented for the calculation of the partition function
$\mathcal{Z}$ of small, isolated and interacting many body
systems. By including anharmonicities and employing a variational
approach quantum effects can be treated even at very low
temperatures. A method is proposed of how  to include collisional
damping. Finally, our approach is applied to the calculation of
the decay rate of metastable systems.
\end{abstract}

\pacs{24.10.Cn, 05.30.-d, 05.60.Gg, 82.20.Xr}

\maketitle


Small many body systems like atomic nuclei and  metal clusters may
undergo self-sustained collective motion. In the former case this
has long been established experimentally for situations where the
nucleons may be assumed to move in a time-dependent state with no
additional excitation energy \cite{bm}. In the last few years one
is also able to measure properties of single, isolated metallic
grains \cite{grains}. Although such systems consist of (many)
particles of identical or similar nature collective motion only
involves one or a few time dependent parameters. The latter must
be introduced so that they obey certain basic relations for the
intrinsic (particle) degrees of freedom. One, formally promising
way is to apply the functional integral technique and to introduce
collective degrees of freedom via a Hubbard-Stratonovich transform
(HST). This is especially useful when the nature of the collective
motion can be guessed and attributed to one generator $\hat{F}$.
Following \cite{bm} one may then introduce a separable two-body
interaction to obtain for the total Hamiltonian
\bel{twobodham}
\hat{H} = \hat{H}_0 + \frac{k}{2} \,\hat{F}\hat{F}\,,
\end{equation}
in which the coupling constant $k$ is negative for iso-scalar
modes. Of course, this might be understood as one prominent term
in an expansion of a general two-body interaction $\hat{V}^{(2)}$
into a complete series of separable terms like $\hat{H} = \hat{T}
+ \hat{V}^{(2)} = \hat{H}_0 + \frac{1}{2} \sum_{\mu=1}^{m} k_{\mu}
\,\hat{F}^{\mu}\hat{F}^{\mu}$. For the sake of simplicity we will
restrict our discussion to the model case (\ref{twobodham}), but
treatments of more terms are feasible and have already been
undertaken for applications which are simpler than those we have
in mind (see e.g. \cite{ath.aly:npa:97}). For zero thermal
excitation the $\hat{H}_0$ might be assumed to simply represent
the dynamics of independent particles. However, in case the
constituents themselves are excited this may no longer be true as
then incoherent, residual interactions
$\hat{V}_{\textrm{res}}^{(2)}$ may play a role, with an ever
increasing role the larger the excitation. In the nuclear case it
is known that single particle excitations acquire a finite width
if they are away from the Fermi level by only $5-8$ MeV
\cite{mahauxsartor:91}. This effect is further enhanced if the
system of particles is thermally excited. In the present Letter we
want to show how these effects may be accounted for in the
functional integral approach. Moreover, in contrast to previous
formulations \cite{ath.aly:npa:97,SPA,PSPA}, we also demonstrate
the usefulness of generalizing the variational formulation of
\cite{var-ital,fer.klh:pra:86} to many body systems. In this way
one may overcome the divergence known to exist for small
temperatures in the case where the collective motion passes over
barrier regions.

Our basic theoretical tool will be the partition function
$\mathcal{Z}$ which at given temperature
$T = 1/(k_{\textrm{B}} \beta)$ may be expressed as
\cite{ath.aly:npa:97}
\bel{partfunc-Fourier} \mathcal{Z}(\beta) =
\sqrt{\frac{\beta}{2\pi|k|}} \int dq_{0} \ \exp
[-\beta\mathcal{F}^\textrm{SPA}(\beta,q_{0})] \ \zeta(\beta,q_{0})
\,.
\end{equation}
The $\mathcal{F}^\textrm{SPA}(\beta,q_{0})$ is a free energy which
depends on the collective variable $q_0$, treated here in the
static limit, which in this context in the literature is referred
to as {\em ``Static Path Approximation''} (SPA) \cite{SPA}. It may
be written as $\mathcal{F}^\textrm{SPA}(\beta,q_{0}) = q_{0}^{2} /
(2|k|) - T \,\textrm{ln} \,z(\beta,q_{0})$, where $z(\beta,q_{0})$
is the grand canonical partition function calculated for the {\em
static} one body hamiltonian
$\hat{\mathcal{H}}_\textrm{HST}[q_{0}] = \hat{H}_0 + q_{0}
\hat{F}$. Actually, the collective variable is "time-dependent"
and is introduced via the HST. Using this manipulation the two
body interaction $k\hat{F}\hat{F}/2$ disappears. The dependence on
the imaginary time $\tau$ is treated through the Fourier series
$q(\tau) = q_{0} + \sum_{r \ne 0} q_{r} \,\exp(\imag\nu_{r}\tau)$,
with the Matsubara frequencies $\nu_{r} = 2\pi r/(\hbar\beta) =
2\pi r T/\hbar$. Genuine quantum effects in collective motion are
hidden in the factor
\bel{corrfactor} \zeta(\beta,q_{0}) = \int \mathcal{D}'q \ \exp
[-s_{\textrm{E}}(\beta,q_{0})/\hbar]\,.
\end{equation}
The key stone for improvements over the SPA is given by the
Euclidean action $s_{\textrm{E}}$. In the so called {\em
``Perturbed SPA''} (PSPA) -- also known as {\em ``SPA+RPA''} or
{\em ``Correlated SPA''} (CSPA) -- this $s_{\textrm{E}}$ is
expanded to second order in the $q_r$. In this way quantum effects
are treated at the level of a local RPA, see e.g.
\cite{PSPA,ath.aly:npa:97,ruc.hoh:pre:01}. In an extension of this
approximation we shall want to make use of terms up to fourth
order and write the action as (see also
\cite{rummel:phd,ruc.anj:epjb:02})
\begin{widetext}
\bel{expandA}
s_{\textrm{E}}(\beta, q_{0}) = \frac{\hbar\beta}{|k|}
\left( \sum_{r,s \ne 0} \lambda_{rs} \,q_{r} q_{s} +
\sum_{r,s,t \ne 0} \rho_{rst} \,q_{r} q_{s} q_{t} +
\sum_{r,s,t,u \ne 0} \sigma_{rstu} \,q_{r} q_{s} q_{t} q_{u} \right) +
\mathcal{O}(q_{r}^{5})
\end{equation}
The interesting point is that the coefficients $\lambda$, $\rho$
and $\sigma$ can be expressed by the one body Green functions
associated with the Hamiltonian
$\hat{\mathcal{H}}_\textrm{HST}[q_{0}]$, which at first may be
assumed to represent simply independent particles for which the
Green function is $g_{k}^{(0)}(z)=(z - \epsilon_{k})^{-1}$. In
such a case the coefficient $\sigma_{rstu}$ of fourth order
consists of terms like \cite{ruc.anj:epjb:02,rummel:phd}
\bal{Cauchy}
& & \frac{|k|}{4!} \sum_{i,k,m,o} F_{io} F_{ki} F_{mk} F_{om} \times \\
& & \left\{ n(\epsilon_{i})
\ g_{o}^{(0)}(\omega_{i} + \imag\nu_{r})
\ g_{k}^{(0)}(\omega_{i} - \imag\nu_{s})
\ g_{m}^{(0)}(\omega_{i} - \imag\nu_{s+t})
+ \ n(\epsilon_{o})
\ g_{i}^{(0)}(\omega_{o} - \imag\nu_{r})
\ g_{k}^{(0)}(\omega_{o} - \imag\nu_{r+s})
\ g_{m}^{(0)}(\omega_{o} - \imag\nu_{r+s+t}) \right.
\nonumber \\
& & \left. + \ n(\epsilon_{k})
\ g_{i}^{(0)}(\omega_{k} + \imag\nu_{s})
\ g_{o}^{(0)}(\omega_{k} + \imag\nu_{r+s})
\ g_{m}^{(0)}(\omega_{k} - \imag\nu_{t})
+ \ n(\epsilon_{m})
\ g_{i}^{(0)}(\omega_{m} + \imag\nu_{s+t})
\ g_{o}^{(0)}(\omega_{m} + \imag\nu_{r+s+t})
\ g_{k}^{(0)}(\omega_{m} + \imag\nu_{t})
\right\} \nonumber
\end{eqnarray}
\end{widetext}
to give just one example of what these quantities look like.

At the level of PSPA it is useful to connect the coefficient
$\lambda$ to the response function $\chi(\omega)$ defined through
$\delta \langle \hat{F} \rangle_{\omega} = -\chi(\omega) \,\delta
q(\omega)$. One gets \cite{ruc.hoh:pre:01}
\bel{lambdachi}
\lambda_{rs} =
\frac{1}{2} \,(1 + k \chi(\imag\nu_{r})) \,\delta_{r,-s} =
\frac{1}{2} \,\lambda_{r} \,\delta_{r,-s} \,.
\end{equation}
The $\lambda_{r}$ serves as the stiffness in $q_{r}$-direction:
\bel{sPSPA}
s_{\textrm{E}}^{\textrm{PSPA}} =
\frac{\hbar\beta}{|k|} \sum_{r>0} \lambda_{r} \,|q_{r}|^{2} =
\frac{\hbar\beta}{|k|} \sum_{r>0}
\frac{\prod_{\mu} (\nu_{r}^{2} + \varpi_{\mu}^{2})}
     {\prod_{k>l}' (\nu_{r}^{2} + \omega_{kl}^{2})} \,|q_{r}|^{2}
\end{equation}
The unperturbed intrinsic excitations $\hbar\omega_{kl} =
\epsilon_{k} - \epsilon_{l}$ are to be calculated from the
eigenvalues of $\hat{\mathcal{H}}_\textrm{HST}[q_{0}]$. The
$\varpi_{\mu}$, on the other hand, are to be found from the
secular equation $1 + k\,\chi(\varpi_{\mu}) = 0$ and represent the
local RPA frequencies. For an unstable collective mode $\mu$ the
$\varpi_{\mu}^{2}$ is negative such that $\lambda_{1}$ may become
negative as well. This happens for temperatures below the so
called {\em crossover temperature} $T_{0}$. There the dynamical
fluctuations in $q_{1}$-direction become too large for the
harmonic approximation to be justified. Formally, this shows up
when in (\ref{corrfactor}) Gaussian integrals become divergent
\cite{ruc.hoh:pre:01,ruc.anj:epjb:02,rummel:phd}. The name
crossover temperature is borrowed from work on "dissipative
tunneling" within the Caldeira-Leggett model (CLM), where a
similar breakdown is observed, see e.g. \cite{weissu}. From this
work it is known that in the crossover region $T \approx T_{0}$
the situation can be cured by treating the anharmonicities in
$q_{1}$-direction explicitly up to fourth order. In this way it is
possible to remove the divergence of the decay rate at $T_{0}$ in
the CLM \cite{grh.olp.weu:prb:87}. In \cite{ruc.anj:epjb:02} this
technique has been applied to many body systems to overcome the
shortcomings of the PSPA. Using this {\em ``Extended PSPA''}
(ePSPA) the partition function can be evaluated down to $T =
T_{0}/2$ where the fluctuations in $q_{2}$-direction become large,
too. At very low temperatures the harmonic treatment of the
fluctuations is no longer justified in {\em any} direction
$q_{r}$. This feature prevents analytic treatments of higher modes
$q_{r}$.

Such problems may largely be abolished by applying a {\em
variational procedure}. In \cite{var-ital,fer.klh:pra:86} the
latter has been developed for the system of a particle moving in a
one dimensional potential. It allows one to evaluate the quantum
$\mathcal{Z}$ at arbitrary temperatures. Using the formalism of
coherent states, these ideas have been applied to many body
systems in \cite{yos.kic.nak.noh:prc:00}. Here we  want to make
use of the expansion (\ref{expandA}). Details of this novel method
will be given in a forthcoming paper. The main idea consists in
rewriting (\ref{corrfactor}) as
\bal{corrfactor-var}
\zeta(\beta, q_{0}) & = &
\int {\cal D}'q \ \exp [-s_{\Omega}^{q_{0}}/\hbar]
\ \exp [-(s_{\textrm{E}} - s_{\Omega}^{q_{0}})/\hbar] \nonumber \\
& = & \zeta_{\Omega}^{q_{0}} \left\langle \exp [-(s_{\textrm{E}} -
s_{\Omega}^{q_{0}})/\hbar] \right\rangle_{\Omega}^{q_{0}}\,.
\end{eqnarray}
The {\em reference action} $s_{\Omega}^{q_{0}}$ is introduced to
specify an averaging procedure for which one may exploit the
inequality
\bel{JensenPeierls}
\left\langle \exp [-(s_{\textrm{E}} -
s_{\Omega}^{q_{0}})/\hbar] \right\rangle_{\Omega}^{q_{0}} \ge \exp
[-\langle s_{\textrm{E}} - s_{\Omega}^{q_{0}}
\rangle_{\Omega}^{q_{0}}/\hbar]
\end{equation}
for an optimization of the expression on the right hand side,
which is easier to evaluate than the average in
(\ref{corrfactor-var}). The reference action should be chosen such
that the normalization factor $\zeta_{\Omega}^{q_{0}} = \int {\cal
D}'q \ \exp [-s_{\Omega}^{q_{0}}/\hbar]$ in (\ref{corrfactor-var})
can be evaluated exactly. A reasonable choice can be constructed
from the PSPA action (\ref{sPSPA}) by replacing the RPA
frequencies $\varpi_{\mu}$ by variational parameters
$\Omega_{\mu}$ \cite{rummel:phd}:
\bel{replace}
s_{\Omega}^{q_{0}} = \frac{\hbar\beta}{|k|} \sum_{r>0}
\frac{\prod_{\mu}  (\nu_{r}^{2} + \Omega_{\mu}^{2})}
     {\prod_{k>l}' (\nu_{r}^{2} + \omega_{kl}^{2})} \,|q_{r}|^{2}
\end{equation}
Evidently, when calculating the integrals for $\langle
s_{\textrm{E}} - s_{\Omega}^{q_{0}} \rangle_{\Omega}^{q_{0}}$ all
terms disappear which are  odd in $q_{r}$. Hence, in the truncated
expansion (\ref{expandA}) only terms of second and fourth order
survive. Using the abbreviation $\Pi_{r} = \prod_{\mu}
(\nu_{r}^{2} + \varpi_{\mu}^{2}) - \prod_{\mu} (\nu_{r}^{2} +
\Omega_{\mu}^{2})$ they can be written as:
\begin{widetext}
\begin{eqnarray}
\langle s_{\textrm{E}}^\textrm{PSPA} - s_{\Omega}^{q_{0}}
\rangle_{\Omega}^{q_{0}}
& = & \hbar \sum_{r>0}
\frac{\Pi_{r}}{\prod_{\mu} (\nu_{r}^{2} + \Omega_{\mu}^{2})}
\label{av2} \\
\langle \delta s_{\textrm{E}}^{(4)} \rangle_{\Omega}^{q_{0}}
& = & \frac{\hbar|k|}{\beta} \sum_{r,s>0} \sigma_{rs-r-s}
\,\frac{\prod_{k>l}' (\nu_{r}^{2} + \omega_{kl}^{2})}
       {\prod_{\mu}  (\nu_{r}^{2} + \Omega_{\mu}^{2})}
\,\frac{\prod_{k>l}' (\nu_{s}^{2} + \omega_{kl}^{2})}
       {\prod_{\mu}  (\nu_{s}^{2} + \Omega_{\mu}^{2})} \ .
\label{av4}
\end{eqnarray}
\end{widetext}
Collecting all contributions the dynamical corrections read
\bel{C-var}  \textrm{ln} \,\zeta^\textrm{var} = \textrm{ln}
\,\zeta_{\Omega}^{q_{0}} - \frac{1}{\hbar} \langle
s_{\textrm{E}}^\textrm{PSPA} - s_{\Omega}^{q_{0}}
\rangle_{\Omega}^{q_{0}} - \frac{1}{\hbar} \langle \delta
s_{\textrm{E}}^{(4)} \rangle_{\Omega}^{q_{0}}\,,
\end{equation}
which cannot be larger than $\textrm{ln}\zeta $. The advantage of
this novel method over the PSPA and the ePSPA is that it can be
applied at any $T$. The reason is that in contrast to the case of
the PSPA  the factors $\nu_{r}^{2} + \Omega_{\mu}^{2}$, which
contribute to the stiffness in $q_{r}$-direction in
(\ref{replace}), stay {\em positive} for all $\Omega_{\mu}$.
Consequently all Gaussian integrals in (\ref{C-var}) converge.

\begin{figure}
\includegraphics[width=65mm]{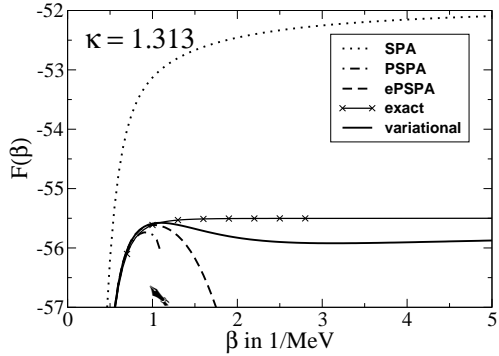}
\caption{\label{fig-Fbeta_lipPBB1313} Free energy of the many body
system for the LMGM with effective coupling constant $\kappa =
1.313$ in various approximations. The arrow points to the inverse
cross-over temperature $\beta_{0} = 1/T_{0}$ where the PSPA breaks
down.}
\end{figure}
To test the results we apply the exactly solvable
Lipkin-Meshkov-Glick model (LMGM) \cite{lih.men.gla:np:65}, as
used before in \cite{ath.aly:npa:97,PSPA} for the the SPA and the
PSPA. In Fig.~\ref{fig-Fbeta_lipPBB1313} the free energy
$\mathcal{F} = -T \ \textrm{ln} \,\mathcal{Z}$ associated to the
total Hamiltonian (\ref{twobodham}) is compared with the
approximations mentioned before. For the calculation the same set
of parameters has been taken as in \cite{ruc.anj:epjb:02}. The
classical SPA gives reasonable results only at small $\beta$ (high
$T$). Inclusion of quantum effects at the level of local RPA
within the PSPA delivers considerable improvement at not too large
$\beta$. However, even before its breakdown at $\beta = \beta_{0}$
it becomes unreliable. The ePSPA behaves completely smoothly in
the crossover region but breaks down at $\beta = 2\beta_{0}$.
Compared to these results, the improvement found for the
variational approach is very striking. Notice please that it is
free of discontinuities and the relative error is only of the
order of 1\%, even at very large $\beta$.

So far the one body Green's functions $g_{k}^{(0)}(z)$ have been
calculated within the picture of independent particle motion. As
mentioned before, within this model the residual two body
interaction $\hat{V}_{\textrm{res}}^{(2)}$ is neglected. This
interaction describes the incoherent scattering of particles and
holes and couples $1$p$1$h states to more complicated ones. This
mechanism may be understood as the origin of the damping of
collective motion, see e.g. \cite{hoh:pr:97}. In the following we
will account for such couplings by dressing the one body Green
functions with self-energies $\Sigma = \Sigma' - \imag\Gamma/2$
meaning that $g_{k}^{(0)}(z)$ is replaced by
\bel{replaceGreen}
g_{k}^{(\Gamma)}(z) =
\left(z - \epsilon_{k} - \Sigma_{k}'(z) +
\imag\Gamma_{k}(z)/2\right)^{-1}
\end{equation}
The dependence of the width $\Gamma_{k}(\omega)$ on frequency and
its variation with temperature is parameterized by the form
\bel{selfimag}
\Gamma_{k}(\omega) = \frac{1}{\Gamma_{0}}
\,\frac{(\hbar\omega - \mu)^{2} + \pi^{2} T^{2}}
{1 + \frac{1}{c^{2}} [(\hbar\omega - \mu)^{2} + \pi^{2} T^{2}]}
\end{equation}
suggested in \cite{siemensetal:84}. For zero temperature it is in
good agreement with empirical data for the widths of proton and
neutron states \cite{mahauxsartor:91}. Finally, some comments are
in order concerning our approximation in handling the impact of
$\hat{V}_{\textrm{res}}^{(2)}$. What is definitely implied in
evaluating many-body Green functions is the application of a
factorization assumption. However, it should be noticed that this
is done for excitations of the intrinsic dynamics and not for the
collective modes. One may therefore argue that coherent effects
are small, in particular at larger thermal excitations. After all
the SPA is meant to represent the high temperature limit.

Having established the connection between the PSPA and the theory
of linear response one may take over the method developed there
(see  \cite{hoh:pr:97}) to extract transport coefficients for
collective dynamics.  For slow collective motion, as given for
nuclear fission, it suffices to concentrate on the lowest mode of
the collective response function. In that frequency regime, the
latter may then be approximated by the response function of a
damped harmonic oscillator, which means the replacement:
\bel{replacechi}
\chi_{\textrm{coll}}(\omega) =
\frac{\chi(\omega)}{1 + k\chi(\omega)}  \quad  \longrightarrow
\quad \chi_{\textrm{osc}}(\omega)
\end{equation}
At any $q_0$ the $\chi_{\textrm{osc}}(\omega)$ is uniquely related
to the coefficients of the local motion such as the inertia,
friction and stiffness. It should be mentioned that exactly at
this point an important difference to the CLM shows up. The latter
only allows a microscopic interpretation of dissipation but not
for the collective potential nor for the inertia. Conversely, our
approach allows us to determine {\em all} coefficients in a
self-consistent fashion, and all of them are influenced by the
microscopic damping mechanism. It may be noted that the
replacement of $g_{k}^{(0)}(z)$ by the $g_{k}^{(\Gamma)}(z)$ also
influences the anharmonic terms of the Euclidean action, see
(\ref{Cauchy}). With respect to the variational approach the
reduction (\ref{replacechi}) leads to a considerable
simplification: Only {\em one} variational parameter $\Omega$ has
to be dealt with in (\ref{replace}) to (\ref{C-var}).

Let us finally turn to the decay rate of meta-stable states. As
shown by J.~S. Langer \cite{laj:ap:69} this rate can be related to
the imaginary part of the free energy. Indeed, the CLM makes
extensive use of this feature \cite{weissu}. We will take this
procedure over to the case of the interacting many body system. In
order to obtain $\textrm{Im} \,\mathcal{F}$ from the $\mathcal{Z}$
in (\ref{partfunc-Fourier}) the integration contour has to be
deformed into the complex $q_{0}$-plane. This has to be done at
the barrier of $\mathcal{F}^{\textrm{SPA}}$ in the sense of a
steepest descent approximation. Details of the resulting rate
formula for interacting many body problems will be given in a
forthcoming paper. The quantum corrections $f_{\textrm{qm}} =
\zeta_{b}/\zeta_{a} \ge 1$ to the purely classical rate formula of
the SPA can be calculated from the dynamical factor $\zeta$ of
(\ref{partfunc-Fourier}) evaluated at the barrier $b$ and at the
minimum $a$ of $\mathcal{F}^{\textrm{SPA}}$ within  the
approximations discussed above.

\begin{figure}
\includegraphics[height=70mm, angle=-90]{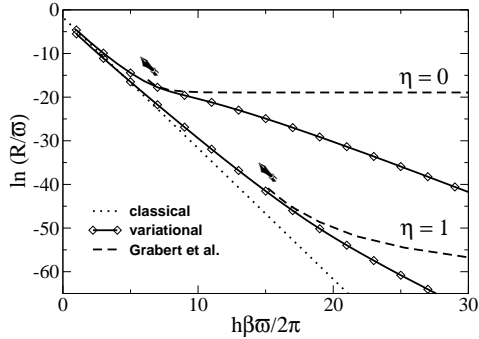}
\caption{\label{fig-rate} A comparison of different approaches to
the decay rate for a particle in a one-dimensional cubic potential
with barrier, for zero ($\eta = 0$) and critical ($\eta = 1$)
damping. The arrows point to the $\beta_{0} = 1/T_{0}$ where the
PSPA breaks down.}
\end{figure}
To compare the various approaches to $\mathcal{Z}$ with respect to
the evaluation of the quantum decay rate  we take the case of a
particle moving in a one dimensional cubic potential with barrier
\cite{grh.olp.weu:prb:87}. For such a situation our variational
approach  reduces to the Feynman-Kleinert method (FKV)
\cite{fer.klh:pra:86}. Results are shown in Fig.~\ref{fig-rate}.
The classical rate (SPA) just displays the exponential behavior of
the Arrhenius law. Quantum corrections are known to increase the
decay rate at larger $\beta$ (low $T$). The variational approach
smoothly matches the classical result to  the low $T$ result
obtained by Grabert et al. with the dynamical ``bounce solution''
\cite{grh.olp.weu:prb:87}. In the region $\beta < \beta_{0}$ the
PSPA would show a similar behavior with the exception of the
crossover region itself. Different to the PSPA the variational
approach gives good results also for $\beta \approx \beta_{0}$ but
also fails in the regime $\beta \gg \beta_{0}$. This behavior may
perhaps be understood as follows. Our variational approach in its
present version starts from a static approximation and only
includes {\em small} quantum corrections. For very large $\beta$
(low $T$) this cannot be enough as one ought to start from genuine
{\em time-dependent} paths $q(\tau)$, like those associated to the
``bounce solution'' in the CLM. For the many body problem with
dynamics beyond independent particle motion, however, this is
still an open problem.

\begin{acknowledgments}
The authors would like to thank J. Ankerhold and F. Ivanyuk for
helpful discussions.
\end{acknowledgments}


\end{document}